\documentclass[prd,twocolumn,preprintnumbers,amsmath,amssymb,superscriptaddress,floatfix,nofootinbib]{revtex4-2}

\usepackage{graphicx} % 插入图片的标准包
\usepackage{dcolumn}  % PRD 标准：让表格中的数字按小数点对齐 (强烈建议保留)
\usepackage{bm}       % PRD 标准：用于书写数学公式中的加粗字体 (比如矢量或矩阵)

\allowdisplaybreaks[4]

\usepackage{hyperref}
\hypersetup{
    colorlinks=true,
    linkcolor=blue,
    anchorcolor=blue,
    citecolor=blue,
    urlcolor=blue % 增加网页链接颜色，方便排版 arXiv 等链接
}

\newcommand{\ucas}{\affiliation{School of Physical Sciences, University of Chinese Academy of Sciences (UCAS), Beijing 100049, China}}
\newcommand{\scnt}{\affiliation{Southern Center for Nuclear-Science Theory, Institute of Modern Physics, Chinese Academy of Sciences, Huizhou 516000, China}}
\newcommand{\thu}{\affiliation{Department of Physics, Tsinghua University, Beijing 100084, China}}

\begin{document}

\title{Investigating nucleon resonances in the $pp\to K^+K^+\Lambda\Lambda$ reaction
}
\author{Zi-Zhao Zhang}
\affiliation{MOE Key Laboratory for Nonequilibrium Synthesis and Modulation of Condensed Matter, School
of Physics, Xi’an Jiaotong University, Xi’an 710049, China.}

\author{Xu Cao}\email{caoxu@impcas.ac.cn}
\affiliation{State Key Laboratory of Heavy Ion Science and Technology, Institute of Modern Physics, Chinese Academy of Sciences, Lanzhou 730000, China}

\author{Bo-Chao Liu}\thanks{\href{mailto:liubc@xjtu.edu.cn}{liubc@xjtu.edu.cn} (corresponding author)}
\affiliation{MOE Key Laboratory for Nonequilibrium Synthesis and Modulation of Condensed Matter, School
of Physics, Xi’an Jiaotong University, Xi’an 710049, China.}

\author{Zi-Xi Ou-Yang}
\ucas

\author{Ke Wang}
\email{wangke1@mail.tsinghua.edu.cn}
\affiliation{Department of Physics and Center for High Energy Physics, Tsinghua University, Beijing 100084, China}

\author{Yin-Jie Wang}
\thu
\author{Jia-Jun Wu}\email{wujiajun@ucas.ac.cn}
\ucas\scnt

\author{Zhi-Hong Ye}\email{yez@tsinghua.edu.cn}
\thu

\begin{abstract}
In this work, we investigate the \(pp \to K^+ K^+ \Lambda \Lambda\) reaction using an effective Lagrangian approach and the isobar model. The calculation includes the exchange of \(\pi\), \(\eta\), \(\sigma\), \(\rho\), \(\omega\), and \(K\) mesons, as well as contributions from eight nucleon resonances: \(N(1535)\), \(N(1650)\), \(N(1710)\), \(N(1720)\), \(N(1895)\), \(N(1900)\), \(N(2060)\) and \(N(2120)\). The double-resonance excitation mechanism (\(NN \to N^* N^*\)) is found to dominate this reaction over the entire energy range considered. Among the exchanged mesons, the \(\eta\) exchange provides the largest contribution at lower energies, whereas the \(K\) exchange dominates at higher energies. Near the threshold, \(N(1535)\) and \(N(1650)\) play the most important roles, while at higher energies \(N(1710)\) and \(N(1720)\) become dominant. Due to isospin conservation, this reaction is predominantly coupled to nucleon resonances, providing a relatively clean channel for \(N^*\) spectroscopy. Our results indicate the unique potential of this reaction for studying nucleon resonances and provide a theoretical reference for future experimental studies at HIAF.
\end{abstract}

\maketitle
\section{INTRODUCTION}
Nucleon resonances ($N^*$), being the excited states of the lightest baryon in quantum chromodynamics (QCD), provide one of the primary probes for investigating the strong interaction in the nonperturbative regime~\cite{Isgur:2000ad}.
At the distance and energy scales relevant to these states, the QCD coupling is large. As a result, two of its most characteristic and least well understood phenomena, namely color confinement and the dynamical generation of hadron mass, are manifest and inherently inaccessible to perturbative methods. Therefore, the excitation spectrum of the nucleon provides a common testing ground where various theoretical approaches can be confronted and cross-checked. 
In this sense, the spectrum offers crucial insights into hadron structure and quark confinement; nevertheless, a complete understanding of it remains a fundamental challenge in hadron physics.

The constituent quark model (CQM), based on SU(6) spin-flavor symmetry, has long provided the benchmark for classifying these excitations. 
It organizes the baryon spectrum into harmonic-oscillator shells, predicting 5, 13, and 30 \(N^*\) states in the \(N=1\), \(N=2\), and \(N=3\) shells, respectively~\cite{Isgur:1978xj,Thomas:1981vc,FlammSchoberl:1982,Burkert:2025coj}. 
While these are model-dependent expectations rather than rigorous QCD predictions, they provide an important reference framework. 
In contrast, experiments have identified only a fraction of the anticipated states. This discrepancy is widely known as the ``missing resonance'' problem~\cite{Anselmino:1992vg,Barabanov:2020jvn}. 
This gap may arise because some states have simply escaped experimental detection, or because the quark model systematically overpredicts the spectrum. 
Resolving which explanation is correct remains a central question in hadron spectroscopy \cite{Doring:2025sgb,Ronchen:2018ury,Lenske:2018bgr,Wang:2017tpe,Mart:2015jof,Cao:2013psa}.

For decades, $N^*$ spectroscopy relied almost exclusively on $\pi N$ scattering, potentially obscuring states that couple only weakly to this channel. 
Over the past two decades, however, high-precision photo- and
electroproduction experiments at ELSA, MAMI, and Jefferson Lab
(CLAS) have led to substantial progress~\cite{Burkert:2025coj, Ireland:2019uwn}. 
By probing a variety of final states and observables, especially
spin-polarization observables, these measurements have provided information complementary to $\pi N$ scattering and key evidence for the discovery or confirmation of numerous resonances~\cite{Anisovich:2017bsk,Burkert:2025coj}. As a result, the lower excitation shells ($N=1$ and $N=2$) are now largely established. However, some candidate $N=2$ states still carry only a one- or two-star rating in the Review of Particle Physics (RPP)~\cite{ParticleDataGroup:2026mpi} (e.g., $N(1860)$, $N(1990)$, and $N(2040)$). Furthermore, several recently proposed states (e.g., $N^{\prime}(1720)$~\cite{Mokeev:2020hhu} and $N(1975)$~\cite{Sarantsev:2025lik}) still need confirmation and are not yet included in the compilation. The situation is particularly acute in the third shell ($N=3$), which is severely underpopulated: the harmonic-oscillator quark model predicts 30 $N^*$ states for this shell, yet only a small fraction has been experimentally confirmed~\cite{ParticleDataGroup:2026mpi,Burkert:2025coj}. The ``missing resonance'' problem therefore persists. Beyond the question of missing states, the masses, widths, and decay branching ratios of observed resonances still carry sizable uncertainties. Moreover, even the internal structure of some well-identified resonances, such as the Roper resonance ($N^*(1440)$), remains actively debated~\cite{Mai:2022eur}: whether the resonance is predominantly a three-quark state or is dynamically generated from coupled channels remains an open question. Further experimental and theoretical efforts are required to resolve these outstanding issues.

To deepen our understanding of nucleon resonances, it is essential to explore the diverse reaction channels through which these states can be excited. 
Because different processes populate resonances with varying strengths, the relative production yields differ considerably.
By selecting appropriate production and decay channels, one can preferentially enhance the contribution of specific resonances---a strategy particularly valuable for states whose production is suppressed or difficult to access via conventional $\pi N$ and $\gamma N$ reactions. 
Such an approach is also crucial for identifying the ``missing'' resonances. 
To this end, nucleon--nucleon ($NN$) scattering offers a complementary and still largely underexploited window~\cite{Chen:2024eaq}.
Indeed, extensive single-meson production studies at COSY have revealed that intermediate baryonic resonances play an important role in the dynamics of exclusive strangeness production in $pp$ collisions~\cite{COSY-TOF:2006tie,TOF:2010ygk,COSY-TOF:2010svd,Wilkin:2016mfn}, demonstrating the potential of $NN$ scattering for probing nucleon resonance properties and refining the resonance spectrum. However, the limited beam energy at COSY restricts the kinematic range accessible for measuring angular distributions, invariant mass spectra, and polarization observables, and data on these observables remain scarce. Such data are essential for identifying the role of nucleon resonances. Furthermore, the energy limit also precludes a systematic exploration of states with masses near and above 2~GeV, whose properties remain poorly constrained.

The recently completed High Intensity Heavy-Ion Accelerator Facility (HIAF) in Huizhou presents a unique opportunity to push these investigations into a significantly higher energy regime~\cite{Yang:2013yeb,Zhou:2022pxl} and offers a rich program for high-precision physics experiments~\cite{An:2025lws}. 
Operating with proton beams up to 10.2~GeV in fixed-target mode, HIAF will reach $pp$ center-of-mass energies of $\sqrt{s}\approx 4.6$~GeV at high luminosity, well beyond the kinematic limits of COSY. 
This substantial increase in beam energy offers distinct advantages in two respects. First, the higher beam energy available at HIAF extends the accessible kinematic domain, enabling systematic measurements of angular distributions, invariant mass spectra, and polarization observables over a broader range; the high beam intensity combined with advanced detector systems at HIAF\cite{Chen:2025ppt,Bai:2026syt} will allow high-luminosity measurements with significantly improved statistical precision. The detailed analyses of these data will help clarify the production mechanisms, thereby improving our understanding of the properties of intermediate resonances. 
Second, the extended kinematic coverage reaches beyond the production thresholds for nucleon resonances above 2~GeV, and also opens access to exclusive multi-strangeness final states, whose underlying dynamics remain largely unexplored. 
HIAF therefore provides a powerful platform for studying the baryon spectrum in a domain where significant open questions remain.

With these opportunities in mind, in the present work we focus on the reaction $pp\to K^+K^+\Lambda\Lambda$, a compelling yet essentially unexplored multi-strangeness channel that will become accessible at HIAF.
The motivation for studying this reaction is twofold. 
On the one hand, for single-meson production in $pp$ scattering, only one nucleon is excited while the other remains in its ground state. 
Within the one-meson-exchange picture, the excitation mechanism involves an $NNM$ vertex on one nucleon line; this mechanism intrinsically suppresses $\eta$- and $\phi$-exchange contributions owing to their exceptionally weak nucleon couplings (e.g., $g_{NN\phi}^2/4\pi \lesssim 0.1$ and $g_{NN\eta}^2/4\pi \approx 0.4$, compared to $g_{NN\pi}^2/4\pi \approx 13.6$)~\cite{Machleidt:2000ge,Rijken:1998yy,Cao:2010km}; this consequently hinders the production of resonances with significant strange-quark components. In contrast, for the double-meson production process considered here, the double-resonance excitation mechanism $NN\to N^*N^*$ is expected to play a dominant role, in which the $NNM$ vertex is replaced by an $N^*NM$ vertex and thereby the $\eta$- and $\phi$-exchange contributions are no longer suppressed. 
Since the ``missing'' $N^{*}$ states may have sizable couplings to strange-quark components, this reaction offers a complementary and potentially more sensitive window into these elusive resonances. 
On the other hand, owing to isospin conservation, the
$K^{+}K^{+}\Lambda\Lambda$ final state couples only to intermediate nucleon resonances, assuming that possible exotic intermediate states such as pentaquarks or tetraquarks can be neglected. 
Although exotic hadrons have been observed experimentally~\cite{BESIII:2013ris,LHCb:2014zfx,LHCb:2015yax,LHCb:2019kea}, no such states relevant to the present reaction have been identified~\cite{ParticleDataGroup:2026mpi}. 
This reaction therefore provides a relatively clean channel for studying nucleon resonances.

In addition, this reaction offers unique and direct access to the
$\Lambda\Lambda$ final-state interaction (FSI). 
The $\Lambda\Lambda$ interaction remains the least-constrained sector of the baryon--baryon force. 
Precise knowledge of this interaction is essential for determining the nature of the $H$-dibaryon~\cite{Jaffe:1976yi}, for resolving the ``hyperon puzzle'' in compact-star physics~\cite{Tolos:2020aln}, and for benchmarking
lattice-QCD predictions~\cite{HALQCD:2019wsz}.
Also, it is possible to search for tetraquarks in $K^+ K^+$ invariant mass spectrum~\cite{Liu:2025fpx,Wang:2024kke,Ji:2024znj}. 
Last but not least, the final-state particles in this process can be efficiently reconstructed and identified.
The $pp\to K^{+}K^{+}\Lambda\Lambda$ reaction at HIAF, with its
high-statistics capability, is ideally suited to provide complementary information on the $\Lambda\Lambda$ interaction and exotic spectrum.
Extracting the $\Lambda\Lambda$ FSI and searching for multiquark states from experimental observables requires a quantitative estimate of how the underlying production mechanism affects the extracted FSI parameters. 
The present work provides a model of the reaction dynamics, which can serve as a starting point for estimating the systematic uncertainties inherent in the FSI extraction procedure and for validating the analysis methodology.
It thus lays the groundwork for subsequent dedicated $\Lambda\Lambda$ FSI investigations in this channel.

The rest of the paper is organized as follows. 
In Sec. II, the theoretical framework and model ingredients are presented. 
In Sec. III, the numerical results are presented and discussed. 
Finally, Sec.~IV provides a short summary.

\section{Theoretical Formalism}

\begin{figure*}[t]
\centering
\includegraphics[width=0.9\textwidth]{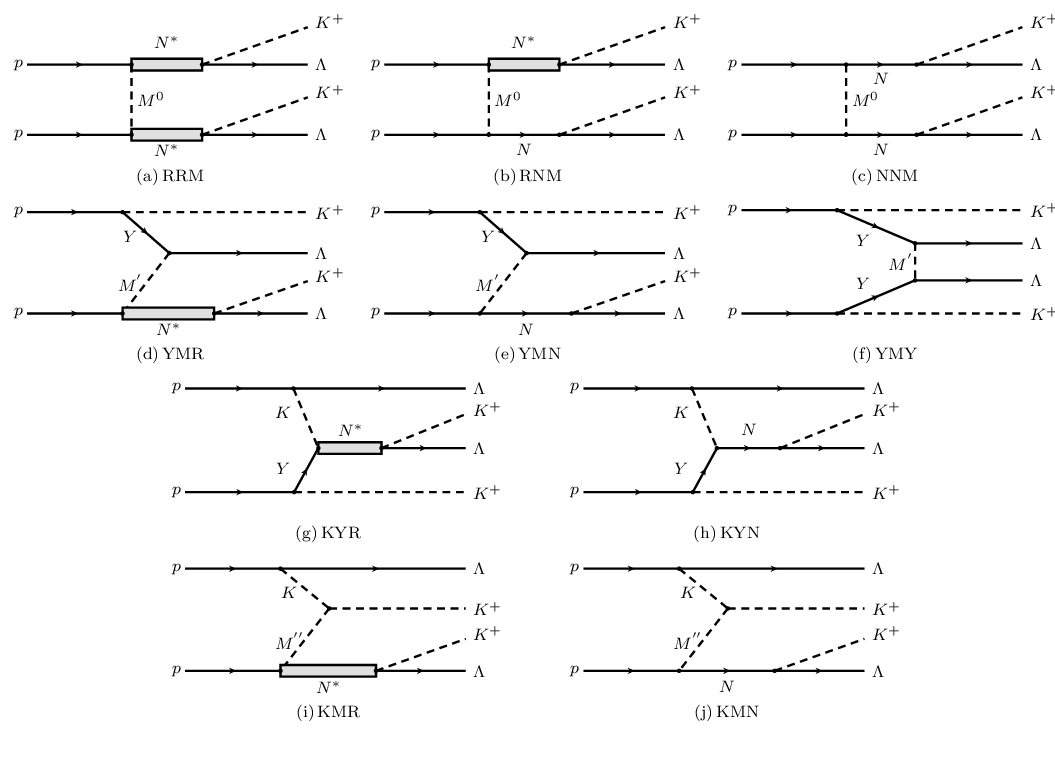}
\caption{Feynman diagrams for the $pp \to K^+ K^+ \Lambda\Lambda$ reaction
considered in this work. The three-letter code after each panel label denotes the internal particles (i.e., the intermediate and exchanged states) of the diagram, with $R$,
$N$, $M$, $Y$, and $K$ representing the nucleon resonance, nucleon, meson,
hyperon ($\Lambda$ or $\Sigma$), and kaon, respectively; these codes serve as
shorthand names for the corresponding diagrams in the following discussion.
The exchanged mesons are specified as $M^0 = \eta, \sigma, \omega, \pi, \rho$;
$M' = \eta, \sigma, \omega$ for $Y = \Lambda$ and $M' = \pi, \rho$ for
$Y = \Sigma$; and $M'' = \sigma, \omega, \rho$.}
\label{feynfig}
\end{figure*}

In this work, we investigate the $pp \to K^+K^+\Lambda\Lambda$ reaction using an effective Lagrangian approach and the isobar model. 
The Feynman diagrams included in our calculation are illustrated in Fig.~\ref{feynfig}. 
For the exchanged mesons, we consider $\pi$, $\eta$, $\sigma$, $\rho$, $\omega$, and $K$ exchanges, while neglecting $K^*$ and $\phi$ exchanges because of their relatively large masses and the limited knowledge of their couplings to nucleon resonances.
For the intermediate nucleon resonances, we include the contributions of the $N(1535)$, $N(1650)$, $N(1710)$, $N(1720)$, $N(1895)$, $N(1900)$, $N(2060)$ and $N(2120)$. 
The $N(1535)$ resonance is included, motivated by its likely significant coupling to the $K\Lambda$ channel~\cite{Yang:2005ej}. 
Other nucleon resonances are taken into account because they have relatively large decay branching ratios to the $K\Lambda$ channel. 
Their corresponding masses, widths, and relevant decay branching ratios are summarized in Table~\ref{tab:resonance_input}. 
For the quantities listed in the table without further specification, we adopt the central values quoted by the RPP~\cite{ParticleDataGroup:2026mpi}. 
For $N(1535)$, we adopt the parameters provided in Ref.~\cite{Liu:2005pm,Liu:2006tf}.

\begin{table*}[htbp]
\centering
\caption{Resonance parameters used in the present calculation. 
Masses and total widths are given in GeV, while branching ratios are given in percent~\cite{ParticleDataGroup:2026mpi}. For the $N(1535)$, we use the parameters of Ref.~\cite{Liu:2005pm}.}
\label{tab:resonance_input}
\begin{tabular}{lcccccccccc}
\hline\hline
State & $M_{N^*}$ & $\Gamma_R$ & $J^P$ & Br$(N\pi)$ & Br$(N\eta)$ & Br$(N\sigma)$ & Br$(N\rho)$ & Br$(N\omega)$ & Br$(\Lambda K)$ & Br$(\Sigma K)$ \\
\hline
$N(1650)$ & 1.650 & 0.125 & $\frac{1}{2}^{-}$ & 60.0 & 25.0 & 10.0 & 17.0 & -    & 10.0 & -    \\
$N(1710)$ & 1.710 & 0.140 & $\frac{1}{2}^{+}$ & 13.0 & 30.0 & 8.0  & 17.0 & 3.0  & 15.0 & -    \\
$N(1720)$ & 1.720 & 0.250 & $\frac{3}{2}^{+}$ & 11.0 & 3.0  & 8.0  & 2.0  & 12.0 & 12.0 & -    \\
$N(1895)$ & 1.895 & 0.120 & $\frac{1}{2}^{-}$ & 10.0 & 30.0 & 7.0  & 32.0 & 28.0 & 13.0 & 13.0 \\
$N(1900)$ & 1.920 & 0.200 & $\frac{3}{2}^{+}$ & 11.0 & 8.0  & 4.0  & 33.0 & 10.0 & 11.0 & 5.0  \\
$N(2060)$ & 2.100 & 0.400 & $\frac{5}{2}^{-}$ & 9.5  & 20.0 & 6.0  & 19.0 & 4.0  & 15.0 & 3.0  \\
$N(2120)$ & 2.120 & 0.300 & $\frac{3}{2}^{-}$ & 10.0 & 3.0  & 9.5  & 1.5  & 12.0 & 8.5  & -    \\
\hline\hline
\end{tabular}
\end{table*}

\subsection{Interaction Lagrangians}
\label{subsec:interaction_lagrangians}

We first introduce the effective interaction Lagrangians employed in this work. 
They are classified according to the spin-parity quantum numbers $J^P$ of the $N^{*}$ states. 
To express the interactions of $N^*$ resonances with the same spin but opposite parity in a compact form, we define $\Gamma^+=1$ and $\Gamma^-=\gamma_5$, which will be used in the following expressions.

For the $N^*$s with $J^P=\frac{1}{2}^{\pm}$, we employ the effective interaction Lagrangians given in Refs.~\cite{Cao:2008etaPrime,Wang:2014ofa,Cao:2007md,Lu:2015phiKLambda,Shyam:1999pKLambda}:
 \begin{align}
\mathcal L_{N^*_{\frac{1}{2}^{\pm}}N\pi}
&=
-i g_{N^*N\pi}\,
\bar{\Psi}_{N^*}\,
\Gamma^{\mp}\,
\boldsymbol{\tau}\cdot\boldsymbol{\pi}\,
\Psi_N
+\mathrm{h.c.},
\\[4pt]
\mathcal L_{N^*_{\frac{1}{2}^{\pm}}N\eta}
&=
-i g_{N^*N\eta}\,
\bar{\Psi}_{N^*}\,
\Gamma^{\mp}\,
\eta\,
\Psi_N
+\mathrm{h.c.},
\\[4pt]
\mathcal L_{N^*_{\frac{1}{2}^{\pm}}N\sigma}
&=
g_{N^*N\sigma}\,
\bar{\Psi}_{N^*}\,
\Gamma^{\pm}\,
\sigma\,
\Psi_N
+\mathrm{h.c.},
\\[4pt]
\mathcal L_{N^*_{\frac{1}{2}^{\pm}}N\rho}
&=
-\frac{g_{N^*N\rho}}{2m_N}\,
\bar{\Psi}_{N^*}\,
\Gamma^{\pm}\sigma_{\mu\nu}\,
\boldsymbol{\tau}\cdot
\left(
\partial^\nu\boldsymbol{\rho}^{\mu}
\right)
\Psi_N
+\mathrm{h.c.},
\label{eq:lag_Nstar12Nrho}
\\[4pt]
\mathcal L_{N^*_{\frac{1}{2}^{\pm}}N\omega}
&=
-\frac{g_{N^*N\omega}}{2m_N}\,
\bar{\Psi}_{N^*}\,
\Gamma^{\pm}\sigma_{\mu\nu}\,
\left(
\partial^\nu\omega^\mu
\right)
\Psi_N
+\mathrm{h.c.},
\label{eq:lag_Nstar12Nomega}
\\[4pt]
\mathcal L_{N^*_{\frac{1}{2}^{\pm}}\Lambda K}
&=
-i g_{N^*\Lambda K}\,
\bar{\Psi}_{N^*}\,
\Gamma^{\mp}\,
K\,
\Psi_\Lambda
+\mathrm{h.c.},\\[4pt]
\mathcal{L}_{N^*_{\frac{1}{2}^{\pm}}\Sigma K} &= -i g_{N^*\Sigma K} \bar{\Psi}_{N^*} \,
\Gamma^{\mp}\,\boldsymbol{\tau}\cdot \boldsymbol{\Psi}_\Sigma K  + \mathrm{h.c.}.
\end{align}

For the $N^*$ states with $J^P=\frac{3}{2}^{\pm}$, described by the Rarita--Schwinger field $\Psi_{N^*}^\mu$, the corresponding interaction Lagrangians are taken from Refs.~\cite{Shyam:1999pKLambda} as

\begin{align}
\mathcal L_{N^*_{\frac{3}{2}^{\pm}}N\pi}
&=
\frac{g_{N^*N\pi}}{m_\pi}\,
\bar{\Psi}_{N^*\mu}\,
\Gamma^{\pm}\,
\boldsymbol{\tau}\cdot
\left(
\partial^\mu\boldsymbol{\pi}
\right)
\Psi_N
+\mathrm{h.c.},
\\[4pt]
\mathcal L_{N^*_{\frac{3}{2}^{\pm}}N\eta}
&=
\frac{g_{N^*N\eta}}{m_\eta}\,
\bar{\Psi}_{N^*\mu}\,
\Gamma^{\pm}\,
\left(
\partial^\mu \eta
\right)
\Psi_N
+\mathrm{h.c.},
\\[4pt]
\mathcal L_{N^*_{\frac{3}{2}^{\pm}}N\sigma}
&=
\frac{g_{N^*N\sigma}}{m_\sigma}\,
\bar{\Psi}_{N^*\mu}\,
\Gamma^{\mp}\,
\left(
\partial^\mu\sigma
\right)
\Psi_N
+\mathrm{h.c.},
\\[4pt]
\mathcal L_{N^*_{\frac{3}{2}^{\pm}}N\rho}
&=
i\frac{g_{N^*N\rho}}{m_{N^*}+m_N}\,
\bar{\Psi}_{N^*\mu}\,
\boldsymbol{\tau}\cdot
\left(
\partial^\nu\boldsymbol{\rho}^{\mu}
-
\partial^\mu\boldsymbol{\rho}^{\nu}
\right)
\nonumber\\[-1mm]
&\qquad\times
\gamma_\nu\Gamma^{\mp}\,
\Psi_N
+\mathrm{h.c.},
\label{eq:lag_Nstar32Nrho}
\\[4pt]
\mathcal L_{N^*_{\frac{3}{2}^{\pm}}N\omega}
&=
i\frac{g_{N^*N\omega}}{m_{N^*}+m_N}\,
\bar{\Psi}_{N^*\mu}\,
\left(
\partial^\nu\omega^\mu
-
\partial^\mu\omega^\nu
\right)
\nonumber\\[-1mm]
&\qquad\times
\gamma_\nu\Gamma^{\mp}\,
\Psi_N
+\mathrm{h.c.},
\label{eq:lag_Nstar32Nomega}
\\[4pt]
\mathcal L_{N^*_{\frac{3}{2}^{\pm}}\Lambda K}
&=
\frac{g_{N^*\Lambda K}}{m_K}\,
\bar{\Psi}_{N^*\mu}\,
\Gamma^{\pm}\,
\left(
\partial^\mu K
\right)
\Psi_\Lambda
+\mathrm{h.c.},
\\[4pt]
\mathcal L_{N^*_{\frac{3}{2}^{\pm}}\Sigma K}
&=
\frac{g_{N^*\Sigma K}}{m_K}\,
\bar{\Psi}_{N^*\mu}\,
\Gamma^{\pm}\,\boldsymbol{\tau}
\cdot 
\boldsymbol{\Psi}_\Sigma \left(
\partial^\mu K
\right)
+\mathrm{h.c.}.
\end{align}

For the $N^*$ states with $J^P=\frac{5}{2}^{\pm}$, described by the rank-two Rarita--Schwinger field $\Psi_{N^*}^{\mu\nu}$, the corresponding interaction Lagrangians are taken from Refs.~\cite{Shklyar:2004spin52,Nakayama:2008eta} as
\begin{align}
\mathcal L_{N^*_{\frac{5}{2}^{\pm}}N\pi}
&=
i\frac{g_{N^*N\pi}}{m_\pi^2}\,
\bar{\Psi}_{N^*\mu\nu}\,
\Gamma^{\mp}
\boldsymbol{\tau}\cdot
\left(
\partial^\mu\partial^\nu\boldsymbol{\pi}
\right)
\Psi_N
+\mathrm{h.c.},
\\[4pt]
\mathcal L_{N^*_{\frac{5}{2}^{\pm}}N\eta}
&=
i\frac{g_{N^*N\eta}}{m_\eta^2}\,
\bar{\Psi}_{N^*\mu\nu}\,
\Gamma^{\mp}
\left(
\partial^\mu\partial^\nu\eta
\right)
\Psi_N
+\mathrm{h.c.},
\\[4pt]
\mathcal L_{N^*_{\frac{5}{2}^{\pm}}N\sigma}
&=
i\frac{g_{N^*N\sigma}}{m_\sigma^2}\,
\bar{\Psi}_{N^*\mu\nu}\,
\Gamma^{\pm}
\left(
\partial^\mu\partial^\nu\sigma
\right)
\Psi_N
+\mathrm{h.c.},
\\[4pt]
\mathcal L_{N^*_{\frac{5}{2}^{\pm}}N\rho}
&=
\frac{g_{N^*N\rho}}{4m_N^2}\,
\bar{\Psi}_{N^*\mu\alpha}\,
\boldsymbol{\tau}\cdot
\partial^\alpha
\left(
\partial^\nu\boldsymbol{\rho}^{\mu}
-
\partial^\mu\boldsymbol{\rho}^{\nu}
\right)
\nonumber\\[-1mm]
&\qquad\times
\Gamma^{\pm}\gamma_\nu\,
\Psi_N
+\mathrm{h.c.},
\label{eq:lag_Nstar52Nrho}
\\[4pt]
\mathcal L_{N^*_{\frac{5}{2}^{\pm}}N\omega}
&=
\frac{g_{N^*N\omega}}{4m_N^2}\,
\bar{\Psi}_{N^*\mu\alpha}\,
\partial^\alpha
\left(
\partial^\nu\omega^\mu
-
\partial^\mu\omega^\nu
\right)
\nonumber\\[-1mm]
&\qquad\times
\Gamma^{\pm}\gamma_\nu\,
\Psi_N
+\mathrm{h.c.},
\label{eq:lag_Nstar52Nomega}
\\[4pt]
\mathcal L_{N^*_{\frac{5}{2}^{\pm}}\Lambda K}
&=
i\frac{g_{N^*\Lambda K}}{m_K^2}\,
\bar{\Psi}_{\Lambda}\,
\Gamma^{\mp}
\left(
\partial_\mu\partial_\nu K^\dagger
\right)
\Psi_{N^*}^{\mu\nu}
+\mathrm{h.c.},
\\[4pt]
\mathcal L_{N^*_{\frac{5}{2}^{\pm}}\Sigma K}
&=
i\frac{g_{N^*\Sigma K}}{m_K^2}\,
\bar{\boldsymbol{\Psi}}_{\Sigma}\cdot
\Big[
\Gamma^{\mp}
\left(
\partial_\mu\partial_\nu K^\dagger
\right)
\boldsymbol{\tau}\,
\Psi_{N^*}^{\mu\nu}
\Big]
+\mathrm{h.c.}.
\end{align}

For the nucleon-pole contribution, we adopt the effective Lagrangians from Refs.~\cite{Cao:2008etaPrime,Lu:2015phiKLambda,Shyam:1999pKLambda}:

\begin{align}
\mathcal L_{NN\pi}
&=
-\frac{f_{NN\pi}}{m_\pi}\,
\bar{\Psi}_N\,
\gamma_5\gamma_\mu\,
\boldsymbol{\tau}\cdot(\partial^\mu\boldsymbol{\pi})\,
\Psi_N,
\\[4pt]
\mathcal L_{NN\eta}
&=
-i g_{NN\eta}\,
\bar{\Psi}_N\,
\gamma_5\,
\eta\,
\Psi_N,
\\[4pt]
\mathcal L_{NN\sigma}
&=
g_{NN\sigma}\,
\bar{\Psi}_N\,
\sigma\,
\Psi_N,
\\[4pt]
\mathcal L_{NN\rho}
&=
-g_{NN\rho}\,
\bar{\Psi}_N
\Biggl[
\gamma_\mu\,
\boldsymbol{\tau}\cdot\boldsymbol{\rho}^{\mu}
\nonumber\\[-1mm]
&\qquad+
\frac{\kappa_\rho}{2m_N}\,
\sigma_{\mu\nu}\,
\boldsymbol{\tau}\cdot
\left(
\partial^\nu\boldsymbol{\rho}^{\mu}
\right)
\Biggr]
\Psi_N,
\\[4pt]
\mathcal L_{NN\omega}
&=
-g_{NN\omega}\,
\bar{\Psi}_N
\Biggl[
\gamma_\mu\,\omega^\mu
+
\frac{\kappa_\omega}{2m_N}\,
\sigma_{\mu\nu}
\left(
\partial^\nu\omega^\mu
\right)
\Biggr]
\Psi_N,
\\[4pt]
\mathcal L_{\Lambda NK}
&=
-i g_{\Lambda NK}\,
\bar{\Psi}_N\,
\gamma_5\,
K\,
\Psi_\Lambda
+\mathrm{h.c.},
\\[4pt]
\mathcal L_{\Sigma NK} &= -i g_{\Sigma NK} \boldsymbol{\bar{\Psi}_\Sigma}\, \gamma_5 \,K^\dagger \boldsymbol{\tau} \Psi_N + \mathrm{h.c.}.
\end{align}

For the hyperon-meson interaction vertices, we adopt~\cite{Zhao:2013LambdaLambda,Xie:2014LambdaSigmaPi,Jiang:2026HyperonDecays}:
\begin{align}
\mathcal L_{\Lambda\Lambda\eta}
&=
-i g_{\Lambda\Lambda\eta}\,
\bar{\Psi}_{\Lambda}\,
\gamma_5\,
\eta\,
\Psi_{\Lambda},
\\[4pt]
\mathcal L_{\Lambda\Lambda\sigma}
&=
g_{\Lambda\Lambda\sigma}\,
\bar{\Psi}_{\Lambda}\,
\sigma\,
\Psi_{\Lambda},
\\[4pt]
\mathcal L_{\Lambda\Sigma\pi}
&=
-\frac{g_{\Lambda\Sigma\pi}}{m_\Lambda+m_\Sigma}\,
\bar{\Psi}_{\Lambda}\,
\gamma_5\gamma_\mu\,
(\partial^\mu\boldsymbol{\pi})\cdot
\boldsymbol{\Psi}_{\Sigma}
+\mathrm{h.c.},
\\[4pt]
\mathcal L_{\Lambda\Lambda\omega}
&=
-g_{\Lambda\Lambda\omega}\,
\bar{\Psi}_{\Lambda}\,
\Bigl[
\gamma_\mu\omega^\mu
+
\frac{\kappa_{\Lambda\Lambda\omega}}{2m_\Lambda}
\sigma_{\mu\nu}(\partial^\nu\omega^\mu)
\Bigr]
\Psi_{\Lambda},
\\[4pt]
\mathcal L_{\Lambda\Sigma\rho}
&=
-g_{\Lambda\Sigma\rho}\,
\bar{\Psi}_{\Lambda}\,
\Bigl[
\gamma_\mu\boldsymbol{\rho}^{\,\mu}
\nonumber\\
&\quad+
\frac{\kappa_{\Lambda\Sigma\rho}}{m_\Lambda+m_\Sigma}
\sigma_{\mu\nu}(\partial^\nu\boldsymbol{\rho}^{\,\mu})
\Bigr]\cdot
\boldsymbol{\Psi}_{\Sigma}
+\mathrm{h.c.}.
\end{align}
For meson interaction vertices, we adopt~\cite{Liu:2012KNEta,Chen:2018OmegaC}
\begin{align}
    \mathcal{L}_{\sigma K K} &= g_{\sigma K K} m_\sigma \sigma K^\dagger K ,\\[4pt]
    \mathcal{L}_{\rho K K} &= i g_{\rho K K} \boldsymbol{\rho}_\mu \cdot \left[ K^\dagger \boldsymbol{\tau} (\partial^\mu K) - (\partial^\mu K^\dagger) \boldsymbol{\tau} K \right], \\[4pt]
    \mathcal{L}_{\omega K K} &= i g_{\omega K K} \omega_\mu \left[ K^\dagger (\partial^\mu K) - (\partial^\mu K^\dagger) K \right]
\end{align}
Here, $\boldsymbol{\tau}$ denotes the vector of Pauli matrices acting in isospin space.

\subsection{Coupling constants}
\label{subsec:coupling_constants}

The coupling constants at the resonance vertices are determined, whenever possible, from the corresponding experimental partial decay widths listed in the RPP~\cite{ParticleDataGroup:2026mpi}:
\begin{equation}
\Gamma_{N^*\to BM}
=
\mathcal B(N^*\to BM)\,\Gamma_{N^*},
\label{eq:partial_decay_width}
\end{equation}
where \(\mathcal B(N^*\to BM)\) is the branching ratio for the decay \(N^*\to BM\), \(\Gamma_{N^*}\) is the total width of the \(N^*\), and \(B\) and \(M\) denote the final-state baryon and meson, respectively.

For a two-body decay \(N^*\to BM\), the magnitude of the three-momentum of the final meson and the energy of the final baryon in the resonance rest frame is given by
\begin{align}
p_M
&=
\frac{
\lambda^{1/2}
\left(
M_{N^*}^{2},m_B^{2},m_M^{2}
\right)
}
{2M_{N^*}},
\label{eq:decay_momentum}
\\
E_B
&=
\frac{
M_{N^*}^{2}+m_B^{2}-m_M^{2}
}
{2M_{N^*}},
\label{eq:decay_baryon_energy}
\end{align}
where
\begin{equation}
\lambda(x,y,z)
=
x^2+y^2+z^2-2xy-2xz-2yz
\label{eq:kallen_function}
\end{equation}
is the K\"all\'en function.

\paragraph{Pseudoscalar-meson vertices.}

The \(\pi\), \(\eta\), and \(K\) mesons can be treated in a
unified manner by denoting the corresponding decay channel as
\(N^*\to BP\), where
\[
(B,P)=(N,\pi),\quad (N,\eta),\quad (\Lambda,K),\quad (\Sigma,K).
\]
Based on the interaction Lagrangians introduced in
Sec.~\ref{subsec:interaction_lagrangians}, the partial widths are
\begin{equation}
\Gamma_{N^*(1/2^\pm)\to BP}
=
C_I\,
\frac{g_{N^*BP}^{\,2}}{4\pi}\,
\frac{p_P}{M_{N^*}}\,
\left(E_B\mp m_B\right),
\label{eq:width_pseudoscalar_12}
\end{equation}
for spin-\(\frac{1}{2}\) resonances, 
\begin{equation}
\Gamma_{N^*(3/2^\pm)\to BP}
=
C_I\,
\frac{g_{N^*BP}^{\,2}}{12\pi}\,
\frac{p_P^3}{m_P^2M_{N^*}}\,
\left(E_B\pm m_B\right),
\label{eq:width_pseudoscalar_32}
\end{equation}
for spin-\(\frac{3}{2}\) resonances and 
\begin{equation}
\Gamma_{N^*(5/2^\pm)\to BP}
=
C_I\,
\frac{g_{N^*BP}^{\,2}}{30\pi}\,
\frac{p_P^5}{m_P^4M_{N^*}}\,
\left(E_B\mp m_B\right).
\label{eq:width_pseudoscalar_52}
\end{equation}
for spin-\(\frac{5}{2}\) resonances.
Here the upper and lower signs correspond to positive- and
negative-parity resonances, respectively. The isospin factor is
taken as
\begin{equation}
C_I=
\begin{cases}
3, & N^*\to N\pi\ \text{or}\ N\rho,\\
1, & N^*\to N\eta,\, N\sigma,\,\Lambda K \,\text{or}\, N\omega\ .
\end{cases}
\label{eq:pseudoscalar_isospin_factor}
\end{equation}

\paragraph{Scalar-meson vertices.}
For the $N^*\to N\sigma$ decay, we have
\begin{equation}
{\Gamma}_{N^*(1/2^\pm)\to N\sigma}
=
\frac{g_{N^*_{1/2}N\sigma}^{\,2}}{4\pi}\,
\frac{p_\sigma}{M_{N^*}}\,
\left(E_N\pm m_N\right)
\label{eq:width_sigma_12}
\end{equation}
for spin-\(\frac{1}{2}\) resonances,
\begin{equation}
{\Gamma}_{N^*(3/2^\pm)\to N\sigma}
=
\frac{g_{N^*_{3/2}N\sigma}^{\,2}}{12\pi}\,
\frac{p_\sigma^3}{m_\sigma^2M_{N^*}}\,
\left(E_N\mp m_N\right)
\label{eq:width_sigma_32}
\end{equation}
for spin-\(\frac{3}{2}\) resonances and
\begin{equation}
{\Gamma}_{N^*(5/2^\pm)\to N\sigma}
=
\frac{g_{N^*_{5/2}N\sigma}^{\,2}}{30\pi}\,
\frac{p_\sigma^5}{m_\sigma^4M_{N^*}}\,
\left(E_N\pm m_N\right)
\label{eq:width_sigma_52}
\end{equation}
for spin-\(\frac{5}{2}\) resonances.

\paragraph{\texorpdfstring{\(\rho\)}{rho}-meson vertices.}

Since some of the $N^*$ resonances considered in this work lie close to the $N\rho$ threshold, the coupling constants of the \(N^*N\rho\) vertices are determined from the cascade decay
\begin{equation}
N^*(p_R)
\to
N(p_1)\rho(q)
\to
N(p_1)\pi(p_2)\pi(p_3)
\label{eq:rho_cascade_decay}
\end{equation} 
with $q=p_2+p_3$, where the symbols in parentheses denote the four-momenta of the corresponding particles. 
The \(N^*_{\frac{1}{2}^\pm}N\rho\) and \(N^*_{\frac{3}{2}^\pm}N\rho\) interactions are described by
Eqs.~\eqref{eq:lag_Nstar12Nrho} and \eqref{eq:lag_Nstar32Nrho}, respectively.
The interaction Lagrangian responsible for the decay
\(\rho\to\pi\pi\) is given by Refs.~\cite{Liu:2012KpKKLambda,Xie:2008NstarNrho}
\begin{equation}
\mathcal L_{\rho\pi\pi}
=
g_{\rho\pi\pi}\,
\left(
\boldsymbol{\Phi}_\pi
\times
\partial_\mu\boldsymbol{\Phi}_\pi
\right)
\cdot
\boldsymbol{\rho}^{\mu}.
\label{eq:lag_rhopipi}
\end{equation}

For spin-\(\frac{1}{2}\) resonances, the cascade amplitude can be
written as
\begin{align}
\mathcal M_{1/2^\pm}^\rho
={}&
\frac{
g_{N^*_{1/2}N\rho}\,
g_{\rho\pi\pi}
}
{2m_N}\,
F_\rho'(q^2)
\nonumber\\
&\times
\bar u_N(p_1)\,
\Gamma^{\pm}\sigma_{\mu\nu}q^\nu\,
u_{N^*}(p_R)\,
G_\rho^{\mu\alpha}(q)
\nonumber\\
&\times
(p_2-p_3)_\alpha .
\label{eq:rho_cascade_amp_12}
\end{align}
For spin-\(\frac{3}{2}\) resonances, it is given by
\begin{align}
\mathcal M_{3/2^\pm}^\rho
={}&
\frac{
g_{N^*_{3/2}N\rho}\,
g_{\rho\pi\pi}
}
{M_{N^*}+m_N}\,
F_\rho'(q^2)
\nonumber\\
&\times
\bar u_N(p_1)\,
\gamma_\nu\Gamma^{\mp}\,
u_{N^*\,\mu}(p_R)
\nonumber\\
&\times
\left(
q^\nu g^{\mu\alpha}
-
q^\mu g^{\nu\alpha}
\right)
G_{\rho,\alpha\beta}(q)
\nonumber\\
&\times
(p_2-p_3)^\beta.
\label{eq:rho_cascade_amp_32}
\end{align}
For spin-\(\frac{5}{2}\) resonances, it is given by
\begin{align}
\mathcal M_{5/2^\pm}^\rho
={}&
-\frac{
g_{N^*_{5/2}N\rho}\,
g_{\rho\pi\pi}
}
{4m_N^2}\,
F_\rho'(q^2)
\nonumber\\
&\times
\bar u_N(p_1)\,
\Gamma^{\pm}\gamma_\nu\,
u_{N^*\,\mu\lambda}(p_R)
\nonumber\\
&\times
q^\lambda
\left(
q^\nu g^{\mu\alpha}
-
q^\mu g^{\nu\alpha}
\right)
G_{\rho,\alpha\beta}(q)
\nonumber\\
&\times
(p_2-p_3)^\beta .
\label{eq:rho_cascade_amp_52}
\end{align}

The $\rho$ propagator $G_\rho^{\mu\nu}$ entering the cascade amplitude is
defined as
\begin{equation}
G_\rho^{\mu\nu}(q)
=
-i\,
\frac{
g^{\mu\nu}
-q^\mu q^\nu/q^2
}{
q^2-m_\rho^2+im_\rho\Gamma_\rho
},
\label{eq:propagator_rho}
\end{equation}
and the corresponding form factor is parametrized as
\begin{equation}
F_\rho'(q^2)
=
\frac{{\Lambda'_\rho}^{2}}{
{\Lambda'_\rho}^{2}+
\left|q^2-m_\rho^2\right|
}.
\label{eq:rho_decay_form_factor}
\end{equation}
To determine the cutoff parameter $\Lambda'_\rho$, we follow the approach of Ref.~\cite{Xie:2008NstarNrho} and relate $\Lambda'_\rho$ to the cutoff parameter used in the $t$-channel $\rho$ exchange through ${\Lambda'_\rho}^{2}+m_\rho^2=\Lambda_\rho^2$. 
With $\Lambda_\rho=1$ GeV adopted in this work, the $\Lambda'_\rho$ is determined to be $0.626$ GeV. 
The $\rho\pi\pi$ coupling constant is taken as $g_{\rho\pi\pi}^2/(4\pi)=2.91$~\cite{Liu:2012KpKKLambda}.

The corresponding three-body partial width is given by
\begin{align}
\Gamma_{N^*\to N\rho\to N\pi\pi}^{\mathrm{th}}
=
\mathcal C_I\,
\frac{1}{2M_{N^*}}
\frac{1}{2J_{N^*}+1}
\int d\Phi_3
\sum_{\mathrm{spins}}|\mathcal M_\rho|^2.
\label{eq:rho_three_body_width}
\end{align}
Here, \(\mathcal C_I\) denotes the isospin coefficient defined in Eq.~(\ref{eq:pseudoscalar_isospin_factor}), and \(d\Phi_3\) is the standard three-body phase space element.
The $N^*N\rho$ coupling constant can then be obtained by setting the theoretical partial width $\Gamma_{N^*\to N\rho\to N\pi\pi}^{\mathrm{th}}$ equal to the corresponding RPP value.

\paragraph{\texorpdfstring{$\omega$}{omega}-meson vertices}
With the interaction Lagrangians introduced above, the partial decay width of a spin-$1/2$ resonance to $N\omega$ is
\begin{equation}
    \begin{aligned}
        \Gamma_{N^*_{1/2^\pm}\to N\omega}
        ={}&
        \frac{g_{RN\omega}^2}{4\pi}
        \frac{p_\omega (E_N\mp m_N)}{4M_{N^*} m_N^2}
        \\
        &\times\left[2(M_{N^*}\pm m_N)^2+m_\omega^2\right].
    \end{aligned}
\end{equation}
For a spin-$3/2$ resonance, the width is
\begin{equation}
    \begin{aligned}
        \Gamma_{N^*_{3/2^\pm}\to N\omega}
        &=
        \frac{g_{RN\omega}^2}{4\pi}
        \frac{p_\omega}{3M_{N^*}}
        \frac{E_N\mp m_N}{(M_{N^*}+m_N)^2}
        \\
        &\times\left[
            2E_N(E_N\pm m_N)
            +(M_{N^*}\pm m_N)^2
            +2m_\omega^2
        \right].
    \end{aligned}
\end{equation}
For a spin-$5/2$ resonance, the width is
\begin{equation}
    \begin{aligned}
        \Gamma_{N^*_{5/2^\pm}\to N\omega}
        &=
        \frac{g_{RN\omega}^2}{4\pi}
        \frac{p_\omega^3}{240M_{N^*}m_N^4}
        \left(E_N\pm m_N\right)
        \\
        &\times\left[
            5m_\omega^2+6p_\omega^2
        \right.
        \\
        &\qquad\left.
            +2(E_N\mp m_N)(E_\omega\mp3m_N)
        \right].
    \end{aligned}
\end{equation}
Here, the upper and lower signs correspond to positive- and negative-parity resonances, respectively. 

The above expressions can be applied to the $N(1895)$ and $N(1900)$ resonances. 
In contrast, the $N(1710)$ and $N(1720)$ lie below or very close to the $N\omega$ threshold, and their finite widths must be taken into account to estimate the couplings to the $N\omega$ channel. 
Following the finite-width prescription of Refs.~\cite{ChenLiu2020,RocaOsetSingh2005,XieLiLiu2020,WangLiu2023}, we use the following equation,
\begin{equation}
    \begin{aligned}
        \Gamma^{\mathrm{FW}}_{N^*\to N\omega}
        ={}&
        -\frac{1}{\pi}
        \int_{s_{th}}^{\infty}
        \mathrm{d}s~\,\Gamma_{N^*\to N\omega}(\sqrt{s})
        F_{N^*}(s)
        \\
        &\times
        \operatorname{Im}\!\left\{
            \frac{1}{s-M_{N^*}^2+iM_{N^*}\Gamma_R}
        \right\},
    \end{aligned}
\end{equation}
to determine the $N^*N\omega$ coupling for the $N(1710)$ and $N(1720)$, where $s_{th}=(m_N+m_\omega)^2$. 
Note that we introduce a form factor $F_{N^*}$ defined in Eq.~(\ref{eq:form_factor_BK}) to regularize the divergence of $\Gamma_{N^*\to N\omega}(\sqrt{s})$ at large $s$.

Using the partial width formulas derived above, together with the total widths and branching ratios provided in RPP~\cite{ParticleDataGroup:2026mpi}, we determine the coupling constants. 
In this work, the central values suggested by the RPP are adopted for all coupling constants except for the \(N(1720)N\omega\) coupling. 
The RPP value for the branching ratio of \(N(1720)\to N\omega\) is \(12\%\sim 40\%\). 
Since the central mass of \(N(1720)\) lies very close to the \(N\omega\) threshold and it couples to the \(N\omega\) channel through \(p\)- and \(f\)-waves, both effects lead to a relatively large \(N(1720)N\omega\) coupling. 
At present, the RPP provides only one entry for this decay, and the \(N(1720)N\omega\) coupling remains poorly determined. 
We therefore adopt the lower limit value suggested by the RPP in our calculation.

The resulting coupling constants are summarized in
Table~\ref{Nscoupling}. 
Since decay widths depend only on the squared coupling constants, the relative signs of the coupling constants cannot be determined by this method. 
In the present calculation, all resonance coupling constants are taken to be positive. 
Note that for the \(N(1535) \Lambda K\) coupling, because the $N(1535)$ lies below the $K\Lambda$ threshold, this constant cannot be extracted from the partial decay width. 
Here we adopt its value from Ref.~\cite{Liu:2005pm}.

For vertices not involving $N^*$ resonances, we take the corresponding coupling constants from the literature. 
The adopted values of the coupling constants, along with their sources, are summarized in Table~\ref{othercoupling}.

%============================================================
\begin{table*}[htbp]
\centering
\caption{
Coupling constants of the nucleon resonances used in the
present calculation.
}

\begingroup
\scriptsize
\setlength{\tabcolsep}{3.2pt}
\renewcommand{\arraystretch}{1.08}
\resizebox{0.6\textwidth}{!}{%
\begin{tabular}{
cccc@{\hspace{1.2em}}cccc
}

\hline

Baryon & $J^P$ & Vertex & $g^2/(4\pi)$
&
Baryon & $J^P$ & Vertex & $g^2/(4\pi)$
\\

\hline

%================================================
% N*(1535) and N*(1650)
%================================================

$N(1535)$
& $\frac{1}{2}^{-}$
& $N^*N\pi$
& $0.035$
&
$N(1650)$
& $\frac{1}{2}^{-}$
& $N^*N\pi$
& $0.037$
\\

&
&
$N^*N\eta$
& $0.281$
&
&
&
$N^*N\eta$
& $0.076$
\\

&
&
$N^*\Lambda K$
& $0.462$
&
&
&
$N^*N\sigma$
& $0.681$
\\

&
&
&
&
&
&
$N^*N\rho$
& $0.946$
\\

&
&
&
&
&
&
$N^*\Lambda K$
& $0.055$
\\

\hline

%================================================
% N*(1710) and N*(1720)
%================================================

$N(1710)$
& $\frac{1}{2}^{+}$
& $N^*N\pi$
& $0.105$
&
$N(1720)$
& $\frac{3}{2}^{+}$
& $N^*N\pi$
& $0.002$
\\

&
&
$N^*N\eta$
& $2.019$
&
&
&
$N^*N\eta$
& $0.079$
\\

&
&
$N^*N\sigma$
& $0.022$
&
&
&
$N^*N\sigma$
& $2.591$
\\

&
&
$N^*N\rho$
& $1.509$
&
&
&
$N^*N\rho$
& $1.755$
\\

&
&
$N^*N\omega$
& $0.319$
&
&
&
$N^*N\omega$
& $11.260$
\\

&
&
$N^*\Lambda K$
& $4.166$
&
&
&
$N^*\Lambda K$
& $0.735$
\\

\hline

%================================================
% N*(1895) and N*(1900)
%================================================

$N(1895)$
& $\frac{1}{2}^{-}$
& $N^*N\pi$
& $0.005$
&
$N(1900)$
& $\frac{3}{2}^{+}$
& $N^*N\pi$
& $0.001$
\\

&
&
$N^*N\eta$
& $0.058$
&
&
&
$N^*N\eta$
& $0.064$
\\

&
&
$N^*N\sigma$
& $0.140$
&
&
&
$N^*N\sigma$
& $0.261$
\\

&
&
$N^*N\rho$
& $0.031$
&
&
&
$N^*N\rho$
& $1.256$
\\

&
&
$N^*N\omega$
& $0.118$
&
&
&
$N^*N\omega$
& $1.772$
\\

&
&
$N^*\Lambda K$
& $0.027$
&
&
&
$N^*\Lambda K$
& $0.109$
\\

&
&
$N^*\Sigma K$
& $0.010$
&
&
&
$N^*\Sigma K$
& $0.024$
\\

\hline

%================================================
% N*(2060) and N*(2120)
%================================================

% N(2060)Npi: 8.584058e-5 rounds to 0.000 at three decimal places.
$N(2060)$
& $\frac{5}{2}^{-}$
& $N^*N\pi$
& $0.0001$
&
$N(2120)$
& $\frac{3}{2}^{-}$
& $N^*N\pi$
& $0.006$
\\

&
&
$N^*N\eta$
& $0.259$
&
&
&
$N^*N\eta$
& $0.162$
\\

&
&
$N^*N\sigma$
& $0.385$
&
&
&
$N^*N\sigma$
& $0.048$
\\

&
&
$N^*N\rho$
& $8.124$
&
&
&
$N^*N\rho$
& $0.020$
\\

&
&
$N^*N\omega$
& $7.800$
&
&
&
$N^*N\omega$
& $0.553$
\\

&
&
$N^*\Lambda K$
& $0.210$
&
&
&
$N^*\Lambda K$
& $0.763$
\\

&
&
$N^*\Sigma K$
& $0.021$
&
&
&
&
\\

\hline

\end{tabular}
}%
\endgroup
\label{Nscoupling}
\end{table*}

\begin{table}[htbp]
\centering
\caption{
Nucleon and additional coupling constants used in
the present calculation. Unless otherwise specified,
all entries denote $g^2/(4\pi)$.
}

\begingroup
\scriptsize
\setlength{\tabcolsep}{4pt}
\renewcommand{\arraystretch}{1.12}

\begin{tabular}{
cc@{\hspace{1.0em}}|@{\hspace{1.0em}}cc
}

\hline\hline

Vertex
& Coupling
&
Vertex
& Coupling
\\

\hline

$NN\pi$
& $f_{NN\pi}^{\,2}/(4\pi)=0.078$\cite{Cao:2010km}
&
$\Lambda\Lambda\eta$
& $4.473$\cite{Zhao:2013LambdaLambda}
\\

$NN\eta$
& $0.400$\cite{Cao:2010km}
&
$\Lambda\Lambda\sigma$
& $3.459$\cite{Zhao:2013LambdaLambda}
\\

$NN\sigma$
& $5.690$\cite{Cao:2010km}
&
$\Lambda\Lambda\omega$
& $8.889\;(\kappa=0)$\cite{Zhao:2013LambdaLambda}
\\

$NN\rho$
& $0.900\;(\kappa=6.1)$\cite{Cao:2010km}
&
$\Lambda\Sigma\pi$
& $6.912$\cite{Xie:2014LambdaSigmaPi}
\\

$NN\omega$
& $20.002\;(\kappa=0)$\cite{Xing:2018axn}
&
$\Lambda\Sigma\rho$
& $0.318\;(\kappa=6.15)$\cite{Jiang:2026HyperonDecays}
\\

$N\Lambda K$
& $14.290$\cite{Kim:2021wov}
&
$\sigma KK$
& $4.241$\cite{Chen:2018OmegaC}
\\

$N\Sigma K$
& $1.210$\cite{Liu:2012KNEta}
&
$\rho KK$
& $0.726$\cite{Liu:2012KNEta}
\\

&
&
$\omega KK$
& $0.726$\cite{Liu:2012KNEta}
\\

\hline\hline

\end{tabular}
\endgroup
\label{othercoupling}
\end{table}
\subsection{Form factors and propagators}
\label{subsec:form_factors_propagators}

To account for the finite size of hadrons and off-shell effects, phenomenological form factors are introduced into the amplitudes.

\paragraph{Form factors.}

For meson exchanges, a form factor is applied at each \(BBM\) vertex with \(B=N,~Y\) or \(N^*\). 
Here $Y$ denotes the hyperon $\Lambda$ or $\Sigma$. 
For the \(NNM\), \(NYK\) or \(YYM\) vertices, the form factors are given by~\cite{Cao:2010km}
\begin{equation}
F_{M}(q^2)
=
\left(
\frac{\Lambda_{M}^2-m_M^2}
     {\Lambda_{M}^2-q^2}
\right)^{n_M},
\label{eq:form_factor_BM}
\end{equation}
with
\begin{equation}
n_M=
\begin{cases}
1, & M=\pi,\eta,\sigma,K\\
2, & M=\rho,\omega.
\end{cases}
\label{eq:form_factor_power}
\end{equation}
Here $q$ denotes the four-momentum of the exchanged meson, and $\Lambda_{M}$ is the corresponding cutoff parameter. 
We take $\Lambda_\pi=\Lambda_\eta=\Lambda_K=1.0$~GeV, $\Lambda_\sigma=1.3$~GeV, $\Lambda_\rho=1.6$~GeV, and $\Lambda_\omega=1.3$~GeV~\cite{Cao:2010km}. 
For the $N^*NM$ or $N^*KY$ vertices with $s$-wave or $p$-wave couplings, we use~\cite{Cao:2010km}
\begin{equation}
F_{M}^*(q^2)
=\frac{\Lambda^{*2}_{M}-m_M^2}
     {\Lambda^{*2}_{M}-q^2}
 \end{equation}
with $\Lambda^*_M=1.0$~GeV for all mesons~\cite{Cao:2010km}. 
For the $d$-wave couplings at the $N(1720)N\sigma$, $N(1900)N\sigma$, $N(2060)N\pi$, $N(2060)N\eta$, $N(2060)N\omega$, $N(2060)N\rho$, $N(2120)N\pi$, and $N(2120)N\eta$ vertices, and for the $f$-wave coupling at the $N(2060)N\sigma$ vertex, we take
\begin{equation}
F_{M}^*(q^2)
=\left (\frac{\Lambda^{*2}_{M}-m_M^2}
     {\Lambda^{*2}_{M}-q^2}\right )^2.
 \end{equation}

For $KKM$ vertices with $M=\rho,\omega,\sigma$, we adopt
\begin{equation}
F'_{M}(q^2)
=
\frac{\Lambda_{M}'^2-m_M^2}
     {\Lambda_{M}'^2-q^2}
\end{equation}
with $\Lambda'_M=1.0$~GeV.

For baryon exchanges, a form factor is introduced for each propagator and is taken as~\cite{Cao:2010km}
\begin{equation}
F_{B}(p^2)
=
\frac{\Lambda_{B}^4}
     {\Lambda_{B}^4+
      \left(p^2-m_B^2\right)^2},
\qquad
B=N,\Lambda,\Sigma,N^*,
\label{eq:form_factor_BK}
\end{equation}
where \(p\) is the four-momentum of the intermediate baryon and \(m_B\) is the mass of the exchanged baryon. 
In this work, we take \(\Lambda_{N}=\Lambda_\Lambda=\Lambda_\Sigma=0.8\)~GeV and $\Lambda_{N^*}=1.0$~GeV~\cite{Cao:2010km}.

\paragraph{Meson propagators.}
The propagator for the exchanged meson with spin 0
is taken as
\begin{equation}
D_M(q)
=
\frac{i}
     {q^2-m_M^2+i\epsilon},
\qquad
M=\pi,\eta,\sigma,K.
\label{eq:propagator_PS_scalar}
\end{equation}

For the exchanged vector mesons, we use
\begin{equation}
D_V^{\alpha\beta}(q)
=
-i\,
\frac{
g^{\alpha\beta}
-q^\alpha q^\beta/q^2
}{
q^2-m_V^2+i\epsilon
},
\qquad
V=\rho,\omega.
\label{eq:propagator_vector}
\end{equation}

\paragraph{Baryon propagators.}
For a spin-\(\frac{1}{2}\) resonance, the propagator is written as
\begin{equation}
S_{N^*}^{(1/2)}(q)
=
\frac{
i\left(\not\! q+M_{N^*}\right)
}{
q^2-M_{N^*}^2
+iM_{N^*}\Gamma_{N^*}
}.
\label{eq:propagator_Nstar_12}
\end{equation}
While the RPP mass and width values are used for other resonances, the $N(1535)$ receives special treatment here because of its possible strong coupling to strangeness channels \cite{Liu:2005pm,Liu:2006tf,Xie:2007qt,Cao:2008etaPrime,Cao:2009ea,Zou:2010tc}. 
To explore its potential contribution in the near-threshold region, we follow Refs.~\cite{Liu:2005pm,Liu:2006tf} and adopt a large $K\Lambda$ coupling with a Flatt\'e form propagator denominator, setting $M_{N^*} = 1400$ MeV and adopting an energy-dependent width:
\begin{equation}
\Gamma_{N^*}(s) = \Gamma^0_{N^*} \big[ 0.8\,\rho_{\pi N}(s) + 2.1\,\rho_{\eta N}(s) + 3.5\,\rho_{\Lambda K}(s) \big],
\end{equation}
where $\Gamma^0_{N^*} = 270$ MeV. 
We have also tested the two-channel parameterization scenario of Ref.~\cite{Liu:2005pm}, retaining only the $\pi N$ and $\eta N$ channels with $M_{N^*} = 1535$ MeV and $\Gamma_{N^*}=150$ MeV. 
The main results are essentially unchanged under this modification. 

For a spin-\(\frac{3}{2}\) resonance, we employ the
Rarita--Schwinger propagator
\begin{equation}
S_{N^*}^{(3/2)\mu\nu}(q)
=
-i\,
\frac{
\left(\not\! q+M_{N^*}\right)
\mathcal P^{\mu\nu}(q)
}{
q^2-M_{N^*}^2
+iM_{N^*}\Gamma_{N^*}
},
\label{eq:propagator_Nstar_32}
\end{equation}
where
\begin{align}
\mathcal P^{\mu\nu}(q)
={}&
g^{\mu\nu}
-\frac{1}{3}\gamma^\mu\gamma^\nu
-\frac{2}{3M_{N^*}^2}\,q^\mu q^\nu
\nonumber\\
&+
\frac{1}{3M_{N^*}}
\left(
q^\mu\gamma^\nu-q^\nu\gamma^\mu
\right).
\label{eq:spin_32_projector}
\end{align}

For a spin-\(\frac{5}{2}\) resonance, we take the
Rarita--Schwinger propagator from Refs.~\cite{Kim:2017Lambda1405,Wang:2026KXi} as
\begin{equation}
S_{N^*}^{(5/2)\mu\nu\alpha\beta}(q)
=
i\,
\frac{
\left(\not\! q+M_{N^*}\right)
\mathcal P^{\mu\nu\alpha\beta}(q)
}{
q^2-M_{N^*}^2
+iM_{N^*}\Gamma_{N^*}
},
\label{eq:propagator_Nstar_52}
\end{equation}
where
\begin{align}
\mathcal P^{\mu\nu\alpha\beta}(q)
={}&
\frac{1}{2}
\left(
\bar g^{\mu\alpha}\bar g^{\nu\beta}
+
\bar g^{\mu\beta}\bar g^{\nu\alpha}
\right)
-
\frac{1}{5}\,
\bar g^{\mu\nu}\bar g^{\alpha\beta}
\nonumber\\
&-
\frac{1}{10}
\Big(
\bar\gamma^\mu\bar\gamma^\alpha\bar g^{\nu\beta}
+
\bar\gamma^\mu\bar\gamma^\beta\bar g^{\nu\alpha}
\nonumber\\
&\qquad+
\bar\gamma^\nu\bar\gamma^\alpha\bar g^{\mu\beta}
+
\bar\gamma^\nu\bar\gamma^\beta\bar g^{\mu\alpha}
\Big),
\label{eq:spin_52_projector}
\end{align}
with
\begin{equation}
\begin{aligned}
\bar g^{\mu\nu}
&=
g^{\mu\nu}
-
\frac{q^\mu q^\nu}{M_{N^*}^2},
\\
\bar\gamma^\mu
&=
\gamma^\mu
-
\frac{q^\mu\not\! q}{M_{N^*}^2}.
\end{aligned}
\label{eq:spin_52_transverse_tensors}
\end{equation}

The propagator for the ground-state baryons is given by
\begin{equation}
S_B(q)
=
\frac{
i\left(\not\! q+m_B\right)
}{
q^2-m_B^2+i\epsilon
}, \qquad B=N,\Lambda,\Sigma.
\label{eq:propagator_nucleon}
\end{equation}

%============================================================
\subsection{Invariant amplitudes and cross sections}
\label{subsec:amplitudes}
%============================================================
Using the model ingredients outlined above, the amplitudes can be constructed via the standard procedure. 
The full amplitude is the coherent sum of contributions from all Feynman diagrams shown in Fig.~\ref{feynfig}, which implicitly include diagrams generated by interchanging the identical particles in the initial or final states. 

The general differential cross section for the $pp\to K^+K^+\Lambda\Lambda$ reaction is given by
\begin{align}
d\sigma =& \frac{1}{64}\, \frac{1}{\sqrt{(p_{B} \cdot p_{T})^2 - m_p^4}}\frac{1}{(2\pi)^8} \sum_{\text{spins}} \left| \mathcal{M}_{fi} \right|^2 \nonumber \\ &\cdot \prod_{a=1}^{4} \frac{d^3 p_a}{2E_a} \; \delta^4(P_i - P_f)\,,
\end{align}
where $\mathcal{M}_{fi}$ is the total amplitude; $P_i$ and $P_f$ denote the total momenta of the initial and final states, respectively; $p_a$ is the momentum of the $a$-th particle in the final state; and $p_B$ and $p_T$ are the beam and target momenta.

\section{Results and Discussion}

In this section, we present our numerical results and discussion for the $pp\to K^+K^+\Lambda\Lambda$ reaction. 
Unless otherwise stated, the total result corresponds to the coherent sum of all contributions from the Feynman diagrams depicted in Fig.~\ref{feynfig}. 
For the differential cross sections, the pure phase-space distribution, normalized to match the integrated total cross section, is also shown for comparison. 
It should be noted that the resonance--baryon--meson coupling constants were obtained mainly from partial decay widths, which do not fix the relative signs. 
Consequently, we simply assume positive signs for these couplings. 
Furthermore, FSI is not included in the present calculation.
Since the primary goal here is to estimate the orders of magnitude of the total cross sections and the roles of nucleon resonances, we regard these approximations as reasonable at the present stage.
% Add references on effective-Lagrangian calculations of nucleon-nucleon induced strange-particle production.

\begin{figure}[t]
\centering
\includegraphics[width=\columnwidth]{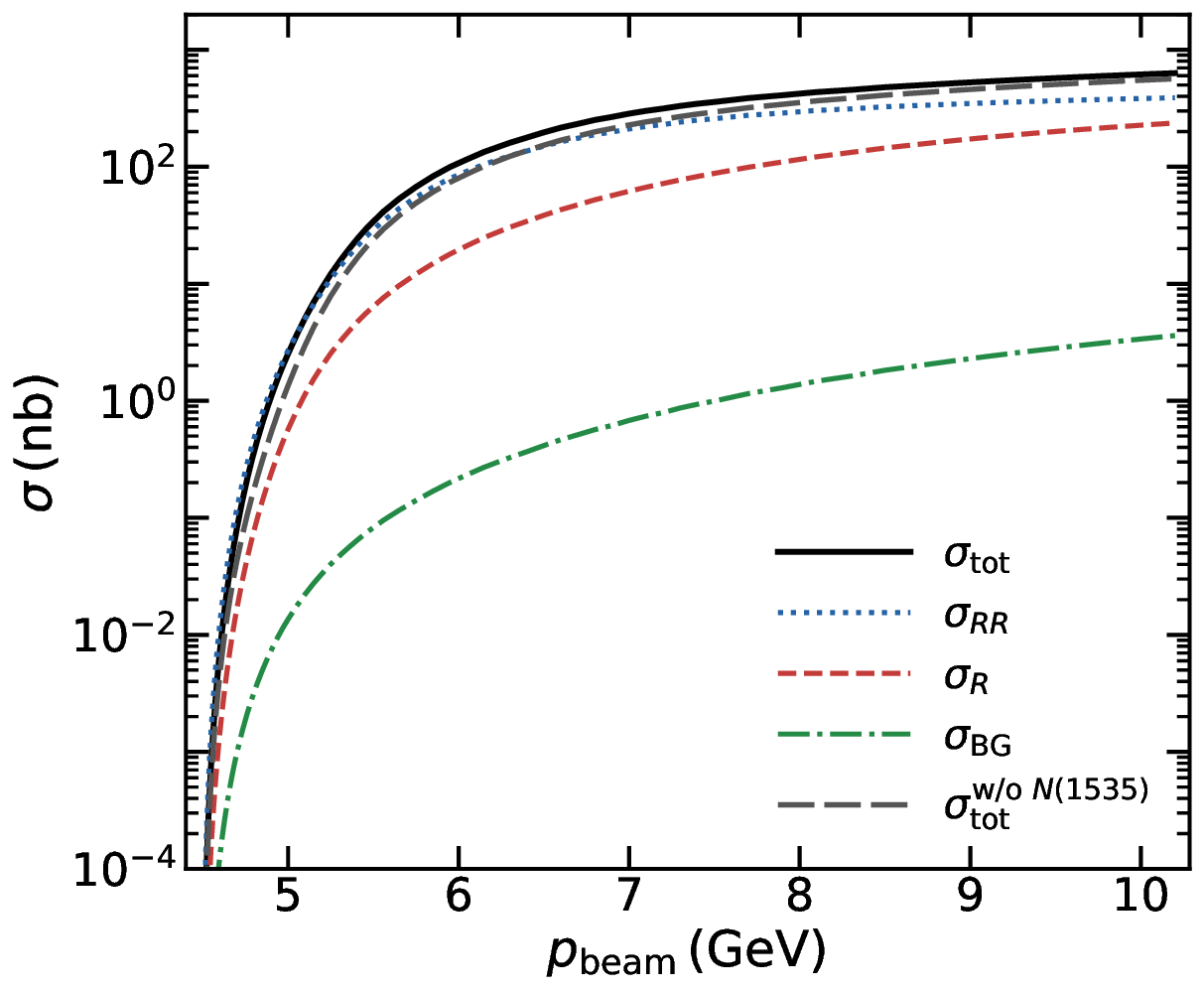}
\caption{Total cross section for the $pp \to K^+ K^+ \Lambda \Lambda$ reaction as a function of beam momentum, along with the individual contributions from the double-resonance, single-resonance, and background mechanisms. The result without the $N(1535)$ contribution is also shown for comparison.}
\label{fig:total}
\end{figure}

\begin{figure}[t]
\centering
\includegraphics[width=\columnwidth]{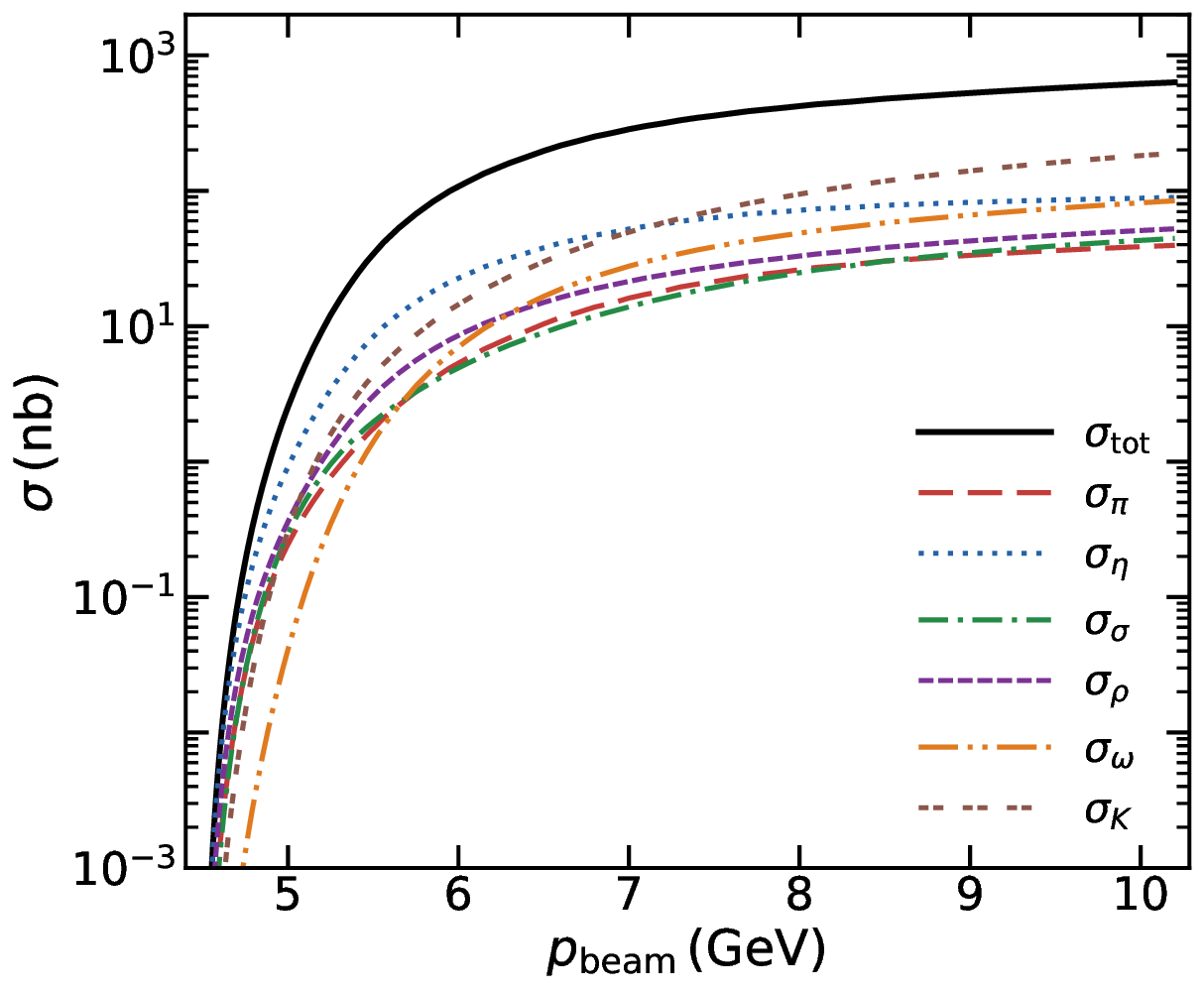}
\caption{Contributions of the $\pi, \eta, \sigma, \rho, \omega$, and $K$ exchange mechanisms to the total cross section. The complete coherent result is shown for comparison.}
\label{fig:meson}
\end{figure}

In Fig.~\ref{fig:total}, we present the total cross section as a function of beam momentum from threshold to 10.2~GeV, together with the individual contributions from the various mechanisms.
Over the entire energy range considered, the contribution from the double-resonance excitation mechanism ($\sigma_{RR}$), corresponding to the (a) $RRM$ diagram denoted in Fig.~\ref{feynfig}, plays the dominant role. 
The $RR$ contribution is almost equal to the total cross section ($\sigma_{total}$) at lower energies, whereas the contribution from the single-resonance excitation mechanism ($\sigma_R$), including the (b) $RNM$, (d) $YMR$, (g) $KYR$, and (i) $KMR$ diagrams, becomes relatively more important as the energy increases. 
The background contributions ($\sigma_{BG}$), including the (c) $NNM$, (e) $YMN$, (f) $YMY$, (h) $KYN$, and (j) $KMN$ diagrams, play only a minor role in this reaction. 
We also show the result obtained by omitting the $N(1535)$ contribution. With the relatively large $N(1535)K\Lambda$ coupling adopted here~\cite{Liu:2005pm}, the $N(1535)$ resonance plays the most important role near threshold (see also Figs.~\ref{fig:ind_Nstar} and~\ref{fig:IM} below). Its relative importance decreases with increasing energy; at 100~MeV above threshold ($p_{\rm beam}=4.84$~GeV), removing the $N(1535)$ contribution still reduces the total cross section by approximately 50\%.

\begin{figure}[t]
\centering
\includegraphics[width=\columnwidth]{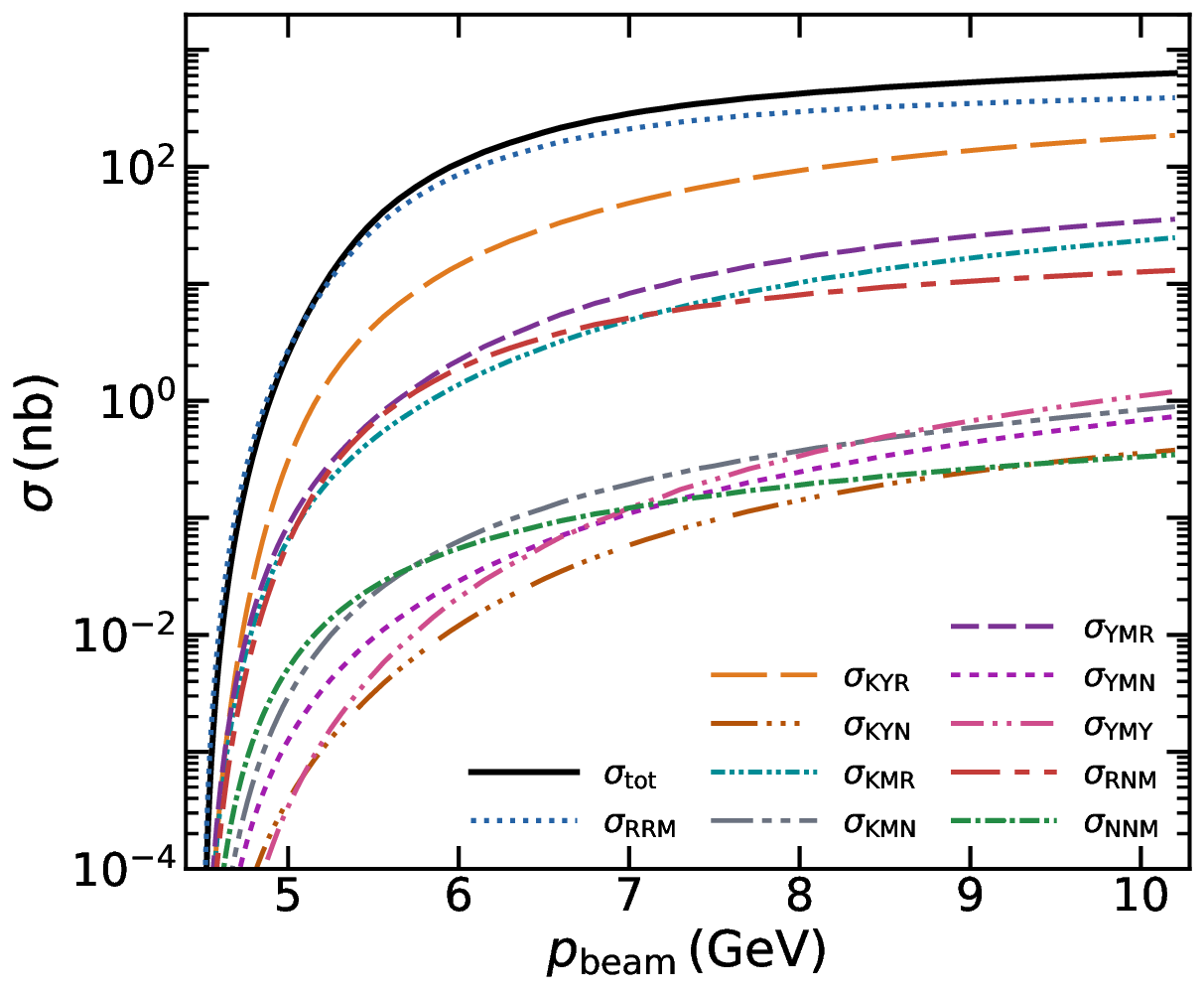}
\caption{Individual contributions from the specific Feynman diagrams (denoted as $\sigma_{RRM}$, $\sigma_{KYR}$, etc., corresponding to those in Fig.~\ref{feynfig}) to the $pp \to K^+ K^+ \Lambda \Lambda$ reaction cross section as a function of beam momentum. The total cross section ($\sigma_{\text{tot}}$) is also shown for comparison.}
\label{fig:individual}
\end{figure}

The individual contributions from the various exchanged mesons are presented in Fig.~\ref{fig:meson}. 
Among the various meson exchanges considered in this work, the $\eta$ and $K$ exchanges yield the largest contributions at lower and higher energies, respectively. 
The contribution from the $\pi$ exchange plays only a relatively minor role in this reaction. 
As discussed in the introduction, the double-resonance excitation mechanism does not suppress the $\eta$ exchange between the initial protons. 
Furthermore, the final $K\Lambda$ channel tends to favor resonances that have large couplings to strange channels, thereby further enhancing the role of the $\eta$ exchange. 
Given that an $\eta$ beam is not available in practice, these results confirm our expectation that the present reaction is suitable for searching for missing resonances. 
The significant $K$ exchange contribution at higher energies is primarily driven by the $KYR$ process, as illustrated in Fig.~\ref{fig:individual}. 
As discussed below, this feature also offers a good opportunity for studying $N^*$ resonances in this reaction.

In Fig.~\ref{fig:individual}, we explore the individual contributions of each Feynman diagram shown in Fig.~\ref{feynfig}; we have implicitly included the contributions from the diagrams obtained by interchanging identical particles in the initial and final states. 
The subscripts on $\sigma$ in the legend denote the corresponding Feynman diagrams and identify the contribution of a specific diagram (see Fig.~\ref{feynfig}). 
As can be seen from the figure, in the single-resonance excitation mechanism, the $KYR$ process plays the most important role. 
The relatively large contribution of this process is mainly due to the large $g_{NKY}$ coupling constant and to the fact that the $N^*K\Lambda$ vertex appears twice in this process, which makes it especially favorable for exciting the $N^*$ resonances that couple strongly to the $K\Lambda$ channel~\cite{Liu:2012KpKKLambda}. 
In the present model, we find that the $N^*(1535)$ and $N^*(1710)$, owing to their large couplings to the $K\Lambda$ channel, give the dominant contributions in the $KYR$ process in the near-threshold and higher-energy regions, respectively. 
This observation shows that, at higher energies, the present reaction via the $KYR$ process offers a unique opportunity to study $N^*$ resonances that have large couplings to the $K\Lambda$ channel and to search for``missing resonance''.

% Add references on baryon-resonance mechanisms in associated strangeness production.

\begin{figure}[t]
\centering
\includegraphics[width=\columnwidth]{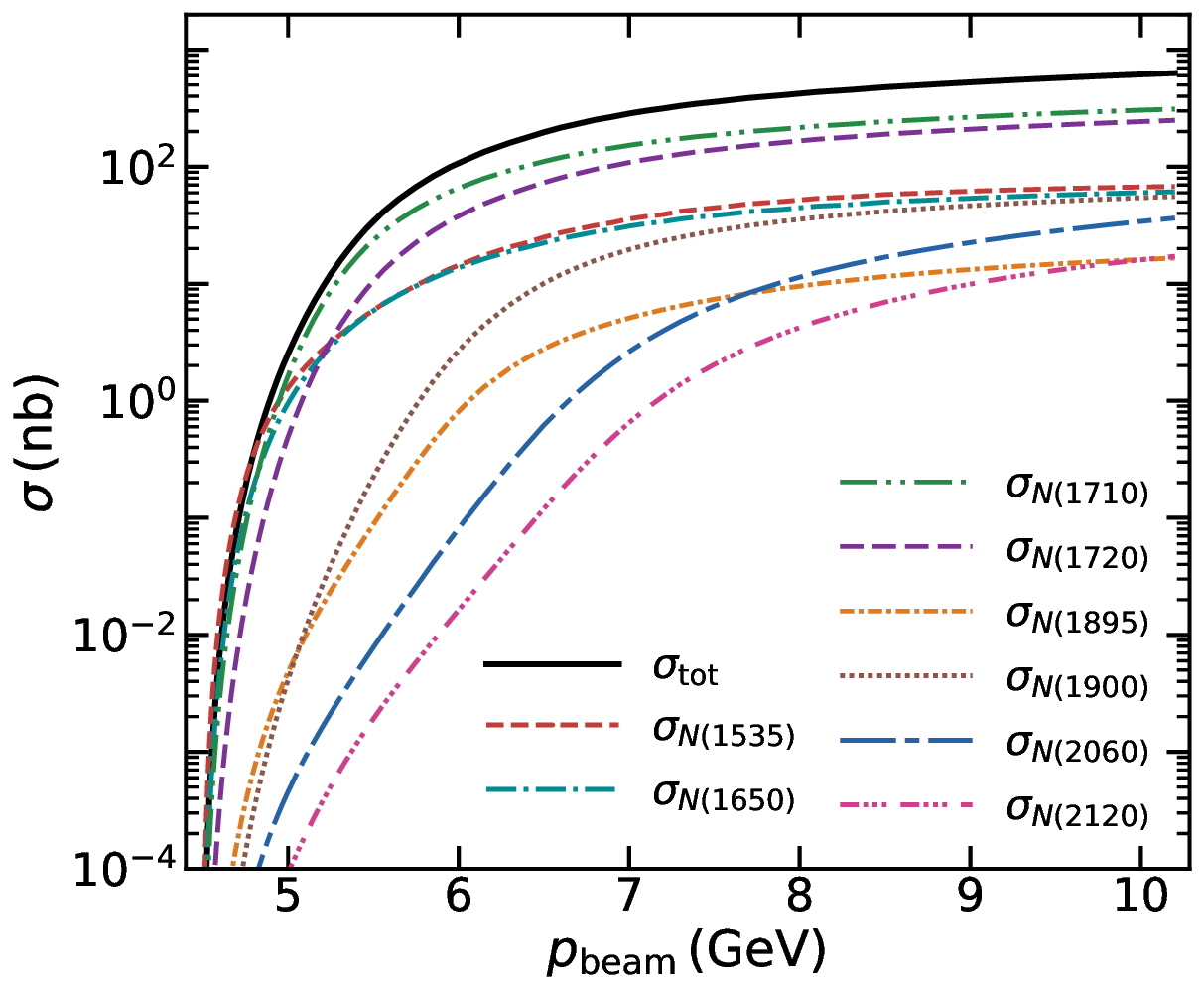}
\caption{Individual contributions of the nucleon resonances to the total cross section as a function of beam momentum. }
\label{fig:ind_Nstar}
\end{figure}

\begin{figure*}[t]
\centering
\includegraphics[width=0.9\textwidth]{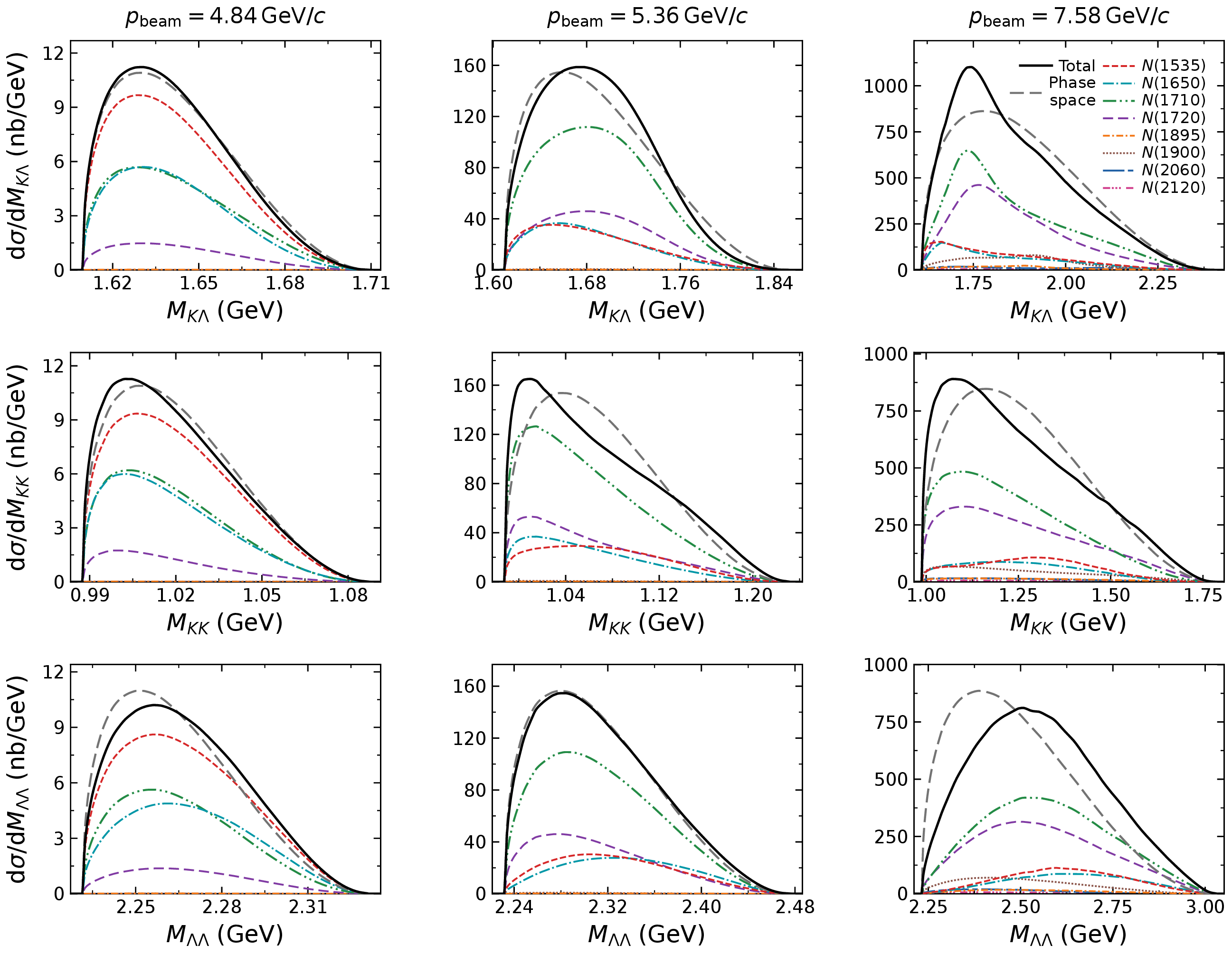}
\caption{Invariant-mass distributions of the two-body subsystems of the final
state, $\mathrm{d}\sigma/\mathrm{d}M_{K\Lambda}$ (top row),
$\mathrm{d}\sigma/\mathrm{d}M_{KK}$ (middle row) and
$\mathrm{d}\sigma/\mathrm{d}M_{\Lambda\Lambda}$ (bottom row), at
$p_{\text{beam}}=4.84$ (left column), 5.36 (middle column) and 7.58 GeV/$c$
(right column). In addition to the complete result (Total), we present the
individual contribution of each nucleon resonance, obtained by including the
contributions of the Feynman diagrams involving the excitation of the
corresponding nucleon resonance. The pure four-body phase-space distribution
is also shown for comparison.}
\label{fig:IM}
\end{figure*}

In Fig.~\ref{fig:ind_Nstar}, we show the individual contributions of each nucleon resonance, obtained by summing the contributions of the Feynman diagrams involving the excitation of the corresponding resonance. Near threshold (for \(p_{\text{beam}} < 4.7\) GeV), the \(N(1535)\) and \(N(1650)\) play the most important roles. 
At higher energies (\(p_{\text{beam}} > 5.0\) GeV), however, the \(N(1710)\) and \(N(1720)\) rise rapidly and become dominant.
Although the \(N(1900)\) contribution is 3-4 orders of magnitude below those of the dominant channels near threshold, it rises steeply with increasing \(p_{\text{beam}}\) and approaches the level of the \(N(1535)\) and \(N(1650)\) contributions above \(p_{\text{beam}} \approx 8\)~GeV; it is therefore no longer negligible at high energies. 
The \(N(1895)\) gives only a minor contribution across the entire energy range, with its cross section about one order of magnitude lower than those of the dominant resonances at high energies and 2--3 orders of magnitude lower near threshold. 
The \(N(2060)\) and \(N(2120)\), due to their relatively large masses, give the weakest contributions over most of the range: near threshold they lie more than three orders of magnitude below the dominant contributions, but they rise steeply
with energy. 
At intermediate energies the \(N(2060)\) gradually surpasses the
\(N(1895)\), and toward the upper end of the considered range both approach the dominant resonances to within about one order of magnitude, with the \(N(2120)\) remaining the smallest contribution throughout.

For the final-state invariant-mass spectra, we focus on the three two-body subsystems, \(M_{K\Lambda}\), \(M_{KK}\), and \(M_{\Lambda\Lambda}\), shown in Fig.~\ref{fig:IM} for the three beam momenta \(p_{\text{beam}}=4.84\), 5.36, and 7.58 GeV/\(c\).
We also present the individual contribution of each nucleon resonance, obtained by summing the contributions of the Feynman diagrams in which the corresponding resonance is excited. The relative importance of the resonances is basically consistent with the cross sections in Fig.~\ref{fig:ind_Nstar}. 
At \(p_{\text{beam}}=4.84\) GeV/\(c\), the \(N(1535)\) gives the most important contribution. 
However, the expected enhancement in \(M_{K\Lambda}\) near threshold does not appear strong, because the phase space available for the \(K\Lambda\) subsystem vanishes at threshold and rises rapidly above it, which masks the near-threshold enhancement. 
With increasing beam momentum, the situation changes: already at \(p_{\text{beam}}=5.36\) GeV/\(c\), the \(N(1710)\) provides the dominant contribution, and at \(p_{\text{beam}}=7.58\) GeV/\(c\), this resonance dominates together with the \(N(1720)\).
Since their masses lie above the \(K\Lambda\) threshold, they can be produced on shell in the \(K\Lambda\) channel, and the corresponding enhancements are clearly seen: at \(p_{\text{beam}}=5.36\) GeV/\(c\), the total \(M_{K\Lambda}\) (\(M_{KK}\)) distribution exceeds the phase-space curve and peaks around 1.68 GeV (1.01 GeV), while at \(p_{\text{beam}}=7.58\) GeV/\(c\), the \(M_{K\Lambda}\) spectrum develops a strong peak around 1.74 GeV with a shoulder near 1.9 GeV, to which the \(N(1895)\) and \(N(1900)\) also contribute, and the \(M_{\Lambda\Lambda}\) peak shifts to about 2.5 GeV, noticeably above the phase-space maximum near 2.4 GeV.

\section{SUMMARY}

We have investigated the $pp\to K^+K^+\Lambda\Lambda$ reaction using an effective Lagrangian approach and the isobar model, motivated by the experimental prospects at HIAF. 
The main findings are as follows.

(1)~The double-resonance excitation mechanism ($NN\to N^*N^*$) dominates this reaction across the entire energy range. 
Because this mechanism does not suppress the $\eta$-exchange contribution, the present reaction is suitable for probing resonances with significant strange-quark components.

(2)~Among the exchanged mesons, the $\eta$ and $K$ exchanges yield the largest contributions at lower and higher energies, respectively. 
Within the single-resonance excitation mechanism, the $KYR$ process plays the most important role.

(3)~Near threshold, the $N(1535)$ and $N(1650)$ dominate. 
At higher energies, the $N(1710)$ and $N(1720)$ play the most important roles. 

These results demonstrate the potential of this reaction for studying nucleon resonances and provide a theoretical reference for future experimental studies at HIAF. 
With future data, our model will also allow a more detailed study of the interference among different amplitudes, making it feasible to investigate the $\Lambda$-$\Lambda$ correlation and quantum entanglement \cite{Zhang:2026nwm,Zhang:2026wvn}, leveraging the self-analyzing property of hyperon decay.

\begin{acknowledgements}
This work is supported by the National Natural Science Foundation of China under Grants No.~12547111 and 12221005,
by the Chinese Academy of Sciences under Grant No. YSBR-101,
and by the Natural Science Basic Research Program of Shaanxi Province under Grants No.~2024JC-YBMS-010 and No.~2026JC-YXQN-004.
Yin-Jie Wang and Zhi-Hong Ye acknowledge support from the Tsinghua University Initiative Scientific Research Program.

\end{acknowledgements}

%以下是参考文献的包
\bibliographystyle{apsrev4-2}  % PRD 推荐的格式
%\bibliography{ref.bib}
\bibliography{ref_cleaned_aps}

\end{document}